\pdfoutput=1
\RequirePackage{fix-cm}
\documentclass[smallextended]{svjour3}       
\smartqed  
\usepackage{subfigure}
\usepackage{graphicx}
\usepackage{wrapfig}
\usepackage{gensymb}
\usepackage{booktabs}  
\usepackage{threeparttable}
\usepackage{multirow}
\usepackage{array}
\usepackage[numbers]{natbib}
\usepackage{amsmath}
\usepackage{authblk}
\usepackage{amsfonts}
\begin{document}
\title{Hepatic vessel segmentation based on 3D swin-transformer with inductive biased multi-head self-attention}


\author[1]{Mian Wu}  
\author[1]{Yinling Qian} 
\author[1]{Xiangyun Liao}
\author[1] {Qiong Wang}
\author[2,1] {Pheng-Ann Heng}
\affil[1]{SIAT, Chinese Academy of Sciences}
\affil[2]{The Chinese University of Hong Kong}

\institute{
}

\authorrunning{Mian Wu \textit{et. al.}}
\titlerunning{Hepatic vessel segmentation based on IBIMHAV-Net}

\date{Received: date / Accepted: date}

\maketitle
\begin{abstract} 
\setlength{\parindent}{0pt} \setlength{\parskip}{1.5ex plus 0.5ex
minus 0.2ex}
\hspace*{\fill} \\ 
\textit{Purpose}: Segmentation of liver vessels from CT images is indispensable prior
to surgical planning and aroused broad range of interests in the medical image analysis
community. Due to the complex structure and low contrast background,
automatic liver vessel segmentation remains particularly challenging. Most of the
related researches adopt FCN, U-net, and V-net variants as a backbone. However, these methods mainly focus on capturing multi-scale local features which may produce misclassified voxels due to the convolutional operator's limited locality reception field. 

\par
\textit{Methods}: We propose a robust end-to-end vessel segmentation network called Inductive BIased Multi-Head Attention Vessel Net(IBIMHAV-Net) by expanding swin transformer to 3D and employing an effective combination of convolution and self-attention. In practice, we introduce the voxel-wise embedding rather than patch-wise embedding to locate precise liver vessel voxels, and adopt multi-scale convolutional operators to gain local spatial information.
On the other hand, we propose the inductive biased multi-head self-attention which learns inductive biased relative positional embedding from initialized absolute position embedding. Based on this, we can gain a more reliable query and key matrix.
To validate the generalization of our model, we test on samples which have different structural complexity.



\par
\textit{Results}: We conducted experiments on the 3DIRCADb datasets. The average dice and sensitivity of the four tested cases were 74.8$\%$ and 77.5$\%$, which exceed results of existing deep learning
methods and improved graph cuts method.

\par
\textit{Conclusion}: The proposed model IBIMHAV-Net
provides an automatic, accurate 3D liver vessel segmentation with an interleaved architecture that better utilizes both global and local spatial features in CT volumes. It can be further extended for other clinical data. 

\keywords  {segmentation \and 3D swin transformer \and multi-head self-attention}

\end{abstract}

\section{Introduction}
\label{intro}

\subsection{Background}
\hspace{1.5em}
CT liver vessel segmentation is essential for 3D visualization, path planning
, and guidance in interventional liver surgery\cite{schumann2012visualization,sboarina2010software}. However, the vessel and liver background show similar intensity values on CT images due to their similarity in the enhancement characteristics. They are curvy, twist, occlude one another, and sometimes are seriously distorted by liver tumors. Due to the intensity similarity and complex structure of the liver vessel, accurate liver vessel segmentation remains some challenge. Nowadays, accurate liver vessel segmentation heavily relies on doctors’ manual segmentation, which is hugely time-consuming and subject to the experience and skills of the experts\cite{chi2010segmentation}. 

Therefore, automatic vessel segmentation has triggered a broad discussion in the community. Even some deep learning methods achieved big success on organ segmentation tasks, they cannot perform well in vessel segmentation due to the considerable variations of vessel structure and unbalance between backgrounds and vessels. Most recent work are designed based on FCN\cite{long2015fully}, U-net\cite{ronneberger2015u}, and V-net's\cite{milletari2016v} variants. They heavily rely on convolution layers, which integrate multi-scale local information to get passable results. Yet convolution's limited reception field does not have long dependencies and enough global features, it can hardly accurately distinguish variant vessel margins and segment minor vessels. 
Therefore, developing a liver vessel segmentation method that adds long dependencies and utilizes global spatial features is necessary. 

\subsection{Related work}
\hspace{1.5em}Current liver vessel segmentation methods can be roughly classified into traditional region-based methods, edge-based segmentation methods and deep learning-based methods. As region-based methods do not perform well in vessel segmentation, we review most related work in the latter two categories. Since we use transformer as our backbone, we also review the newest work related to transformer. A more comprehensive literature survey can refer to \cite{ciecholewski2021computational}. 

\hspace*{\fill} \\
\emph{Traditional methods} \\
Edge-based methods can be further classified into image filtering and enhancement algorithms, tracking-based algorithms \cite{moccia2018blood}. 
Filter and enhancement algorithms extract the volume with a common process called filtering to reduce the noise, then enhance the vessels by applying image gradients or multi-scale high order deviations, particular the second derivatives of the angiographic images to extract high-frequency information\cite{luu2015quantitative,lamy2021vesselness}. Besides, \citet{pamulapati2011intra} introduced a vessel segmentation method based on the medial axis enhancement filter. Tracking-based algorithms focus on the predefined vessel models and track the minimum cost path. \citet{friman2010multiple} proposed to track many hypothetical vessel trajectories at the same time, which improved the results in low contrast conditions. \citet{cetin2012vessel,cetin2015higher} presented the tubular structure segmentation method, which utilized a second-order tensor from directional intensity measurement and employed higher-order tensor based on cylindrical flux-based to construct the vascular structure.
 
\hspace*{\fill} \\
\emph{Deep learning-based methods} \\
Most deep learning-based liver vessel segmentation work rely on CNN-based architecture, specifically, U-net\cite{ronneberger2015u} and it variants, as well as little attempts by FCN\cite{long2015fully} and V-net\cite{milletari2016v}. In chronological order, early stage vessel segmentation methods like retinal vessel segmentation based on 2D methods. Later, with the segmentation targets changed to 3D images, 3D methods became the mainstream.   
\citet{fu2016deepvessel,li2015cross} have proposed the segmentation method for the retinal vessel from 2D images. These methods can handle small objects in 2D slice, howeve the vessel segmentation on liver, brain, or lung are volume tasks. Most 2D methods cannot transfer to 3D images directly due to space continuous along the Z-axis, which omits essential information. Therefore, the current state of art solutions for liver vessel segmentation focus on 2D multi-path(2.5d) and 3D methods. \citet{kitrungrotsakul2019vesselnet} specifically proposed three DenseNets with shared kernel that fit for resampling three planes(sagittal, coronal and transverse planes) patches from IRCADb dataset called 2.5D method. \citet{cciccek20163d} extend UNet from 2D image to volume, which fused multi-scale 3D convolution feature called 3D-UNet.  
In order to employ the 3D representation of liver vessel features, \citet{huang2018robust} proposed the variant of 3D-Unet fit the problem worked well, their evaluation on IRCADb incomplete annotations further improved result. \citet{yu2019liver} added the residual module into the 3D-UNet that provided more residual features. \citet{xu2020training} employed a 3D-FCN frame for this task. However, a reasonable supervised deep network model has to be trained on a large dataset with high-quality labels, and the current datasets cause the noise labels to hurt the model performance.
Lately, \citet{yan2020attention} proposed a way to fuse self-attention into 3D U-net that improved segmentation details as a great attempt.
\hspace*{\fill} \\
\emph{Vision transformers and 2D swin transformer} \\
 The self-attention mechanism allows transformers to dynamically extract the important features of word sequences and learn their long-range dependencies. This notion has recently been extended to computer vision by defining the vision transformer\cite{dosovitskiy2020image}, which aims at the image recognition task. By taking 2D image patches with positional embeddings as input and pre-trained on large classical dataset, ViT achieved comparable results with the CNN-based methods. In medical image tasks, more recent methods like \cite{zhang2021transfuse,chen2021transunet} enjoyed the benefit of both CNNs and transformers. Effort of \citet{chen2021transunet} firstly utilize CNNs to extract low-level local features and transformers to catch global intersections. Currently, based on the shifted windows mechanism, \citet{liu2021swin} proposed Swin transformer that can learn hierarchical object concepts at different scales by applying appropriate downsampling to feature maps that achieved state-of-art semantic segmentation.
Inspired by swin-transformer, Swin-Unet\cite{cao2021swin} firstly employed hierarchical transformer blocks with integrated encoder and decoder to build U-shape architecture. This work improved transUnet's result on medical multi-organ segmentation tasks. For 3D segmentation, \citet{karimi2021convolution} tentatively replace the 3D convolutional operators with transformers as the backbone to build the model. They first split the local volume block into 3D patches and embedded them into 1D sequence and through ViT's self-attention design. Compared to these methods, our IBIMHAV-Net inherits advantages of convolution in encoding precise spatial information and using inductive biased self-attention in hierarchical representation that helps to overcome connectivity and variance of liver vessel segmentation.


\subsection{Proposed method}
\hspace{1.5em}
Motivated by existing 2D swin-transformer\cite{cao2021swin,liu2021swin} and past vision transformer attempts\cite{chen2021transunet, dosovitskiy2020image,hatamizadeh2021unetr}, we propose a transformer-based architecture for volumetric liver vessel segmentation which better utilize global features and long dependencies.   
The main advantages and contributions of the proposed method are as follows:

1. We propose a network architecture by expanding swin transformer to 3D and combining convolution and self-attention to play their strengths. For self-attention, the global spatial information has been encoded by embedding, and long dependencies have been entangled by our designed 3D transformer block. For convolution, multi-scale convolutions in the local feature path and downsampling/upsampling layers help to encode precise local information and capture hierarchical resolution features. 

2. We introduce the voxel-wise rather than patch-wise embedding as the initial transformer input to fully utilize volumetric information, which transforms volumetric prediction to the sequence to sequence prediction in hierarchical resolution features.

3. We propose the Inductive Biased multi-head attention(IB-MSA) which changes the positional embedding way that learns biased positional embedding with an initialization of absolute 1-dimensional embedding in the transformer blocks.
Thus dramatically improves liver vessel segmentation results. 
\par

\section{Methodology}
\label{sec:1}
 The proposed method starts with dataset preprocessing. Then we introduce architecture of our framework, namely Inductive BIased Multi-Head Attention Vessel Net(IBIMHAV-Net), including the details of our 3D transformer design and inductive biased multi-head attention mechanism. Finally, we describe post-processing which reduces some discrete inaccurate results. 

\subsection{Preprocessing}
Preprocessing plays an essential role and affects the segmentation results significantly\cite{huang2018robust,isensee2018nnu}. For example, applying preprocessing to lower the background noises and augment image contrast. Therefore, we arranges preprocessing as 4 steps: 
(1)3D IRCADb provides 20 groups of CT images, liver vessel masks and liver masks. We crop CT images and liver vessel masks to liver region boundary as the ROI. Then adjust to the size to 256x256x192 to unify the model input. 
(2)We truncate the intensity of all voxels in the volumes to the
range of [-50, 250] HU to reduce the irrelevant details and
increase image contrast. (3)In order to remain enough vessels' continuity features, we add vessel mask outside the liver as supplement of vessel information(Eg.\textbf{Fig.1}). (4)Images are normalized to zero mean and unit variance. Because most liver vessels are quite small, we keep images with their original resolution can prevent artifact errors caused by resampling. 

\begin{figure}
  \centering
  \includegraphics[scale=0.2]{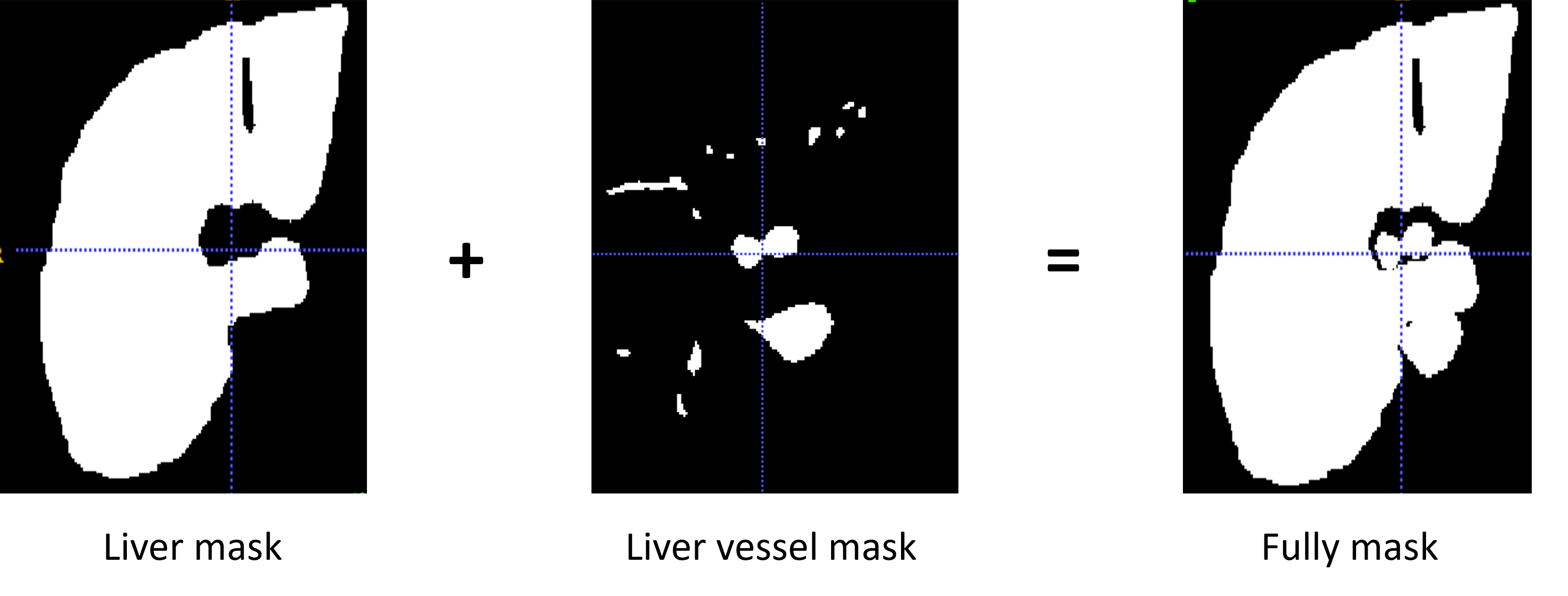}
  \label{fig1}
  \caption{supplement of vessel mask used in the training set}
\end{figure}

\subsection{Overview of the architecture}
The overall proposed architecture is shown in \textbf{Fig.2} Left, which illustrates its U-shape form which includes encoder and decoder. We introduce the U-shape end-to-end Transformer network IBIMHAV-Net, which employ pixel-wise embedding way for transformers. Our model's long-range contextual interactions and precise spatial locate dependencies was provided by inductive biased multi-head self attention(IB-MSA) modules. The U-shape structure combined with feature extract path and three skip connections between multi-scale feature pyramids of encoder and decoder in a symmetrical manner. It helps to keep fine-grained details between transformer blocks. The feature extraction block and interleaved convolution up/downsampling layers gain accurate local spatial information and abundant local features. 

\begin{figure}
  \centering 
  \includegraphics[height=120mm,width=105mm]{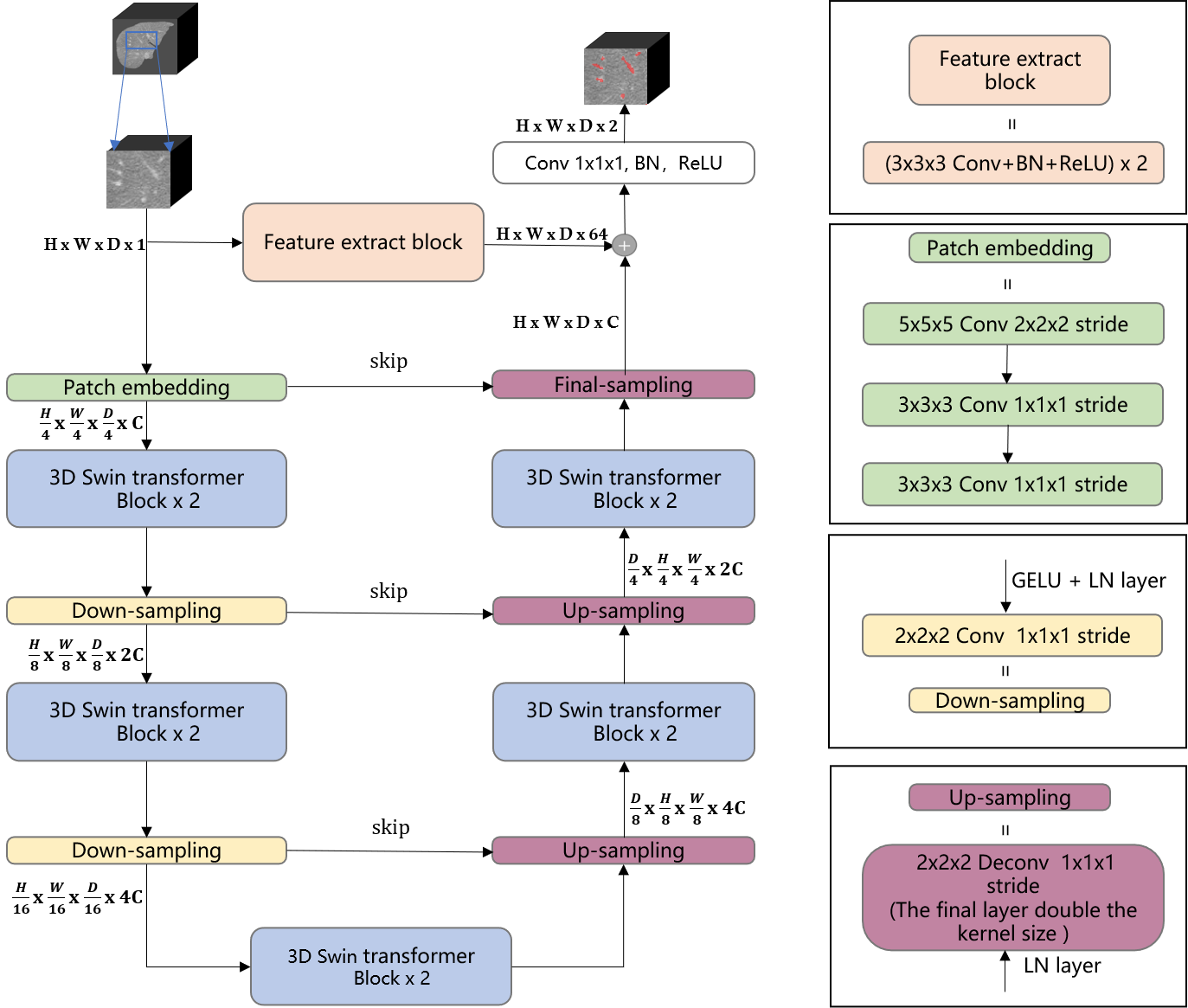}
  \caption{Left: Architecture of IBIMHAV-Net. 
           Right: Compose of conv embedding, feature extraction block, up-sampling layers and down-sampling layers. }
\end{figure}

\subsection{Encoder} \label{encoder set}
Past vision transformer work like\cite{liu2021swin,dosovitskiy2020image,chen2021transunet} have complete encoder part, yet they did not design a 3D encoder.
Our architecture build up a path that includes 3D embedding block, downsampling layers combine with our transformer block design.  
In the encoder, the input is a 3D volume patch randomly cropped from full volume. Then we represent each 3D patch as HxWxD where H,W,D denote the height, width and depth of each input patch, respectively. Thus, the 3D convolution embedding layer obtains tokens, with each patch/token consisting of a 128-dimensional feature. A linear embedding layer is then applied to project the features of each token to a 1D sequence length denoted by C. The outputs of patch embedding block are connected to five 3D swin transformer blocks interspersed with down-sampling blocks.  

\hspace*{\fill} \\
\noindent \textbf{The  patch embedding block}
 The linear embedding part is essential in the original swin transformer design\cite{liu2021swin}, the Swin-T version first splits the one channel vessel patch into non-overlapping patches size of into 1-D sequence, then followed with big convolutional kernels in the linear embedding layer to extract small patches features.
However, our task needs more precise spatial information with larger input volume. Our embedding layer first tokenized  the vessel volume patch $\mathcal{X} \in \mathcal{R}^{H \times W \times D}$ into  high dimensional tensor. This high-dimensional tensor represents as $\mathcal{T} \in \mathrm{R}^{\frac{H}{4} \times \frac{W}{4} \times \frac{D}{4} \times C}$, where ${\frac{W}{4} \times \frac{D}{4} \times C}$ is the patch tokens and C represents the length of sequence which is 128(discussed in \ref{Section:compare}). Due to the variant and complex vessel structure, we design the successive large kernel convolutional combinations for pixel-wise level sequence encoding instead patch-size encoding. Moreover, this setting reduce computational complexity with same range of receptive field to accommodate long sequence. After every convolutional layer followed one GELU and one layerNorm layer to fully embedding as 1-D sequence. The kernels and strides are set as \textbf{Fig.2} Right since the input volumes were nearly squares to fit the model.   

\hspace*{\fill} \\
\noindent \textbf{Down-sampling layer}
The swin transformer blocks used neighboring concatenate operation in past 2D tasks\cite{cao2021swin,liu2021swin}. However, we find that easy convolution with small strides worked better. It also needs a GELU layer and a Layer Norm to keep normalization of processing measures to refine the feature map mapped to [0, 1] to keep the sensitivity of model. It works better than Batch Nomalization(BN) and ReLU activation function in our architecture. 

\subsection{3D swin transformer block with Inductive Biased MSA Module}
After passing patch embedding block's, the high dimensional sequence tensor $\mathcal{T} $ is put into transformer blocks. 
Compare to original Swin transformer, our method conduct self-attention in a hierarchical path and compute self-attention within 3D patches volume with bias focusing on block edge segmentation (i.e. IB-MSA, bias positional multi-head self-attention) instead 2D shift window. 

\hspace*{\fill} \\
\noindent \textbf{3D transformer block}
In the tail of embedding block, the sequence is transformed to the high-dimensional tensor in swin transformer blocks. The main idea is to fully mix the captured long-term dependencies with hierarchical object concepts at various scales with following down-sampling convolution and global spatial information from beginning embedding block. 

In order to represent the workflow in our design, let the high-dimensional tensor $\mathcal{T} \in \mathcal{R}^{L \times C}$ reshape as $\hat{\mathcal{T}} \in \mathbf{R}^{N \times P \times C}$ by passing through IB-MSA, where $N$ is the number of tiny local volumes, $P = S_{H} \times S_{W} \times S_{D}$ denotes the number of patch tokens in each volume. $\left\{S_{H}, S_{W}, S_{D}\right\}$ stand for the size of tiny local volume. To fit to our task's
various shape of vessel CT scans, this setting could cover all patch tokens of last transformer block in encoder.
Because different sampling quality between datasets, it may not be reasonable to brute-forcely pad the data in order to satisfy fixed $\left\{S_{H}, S_{W}, S_{D}\right\}$. Therefore, the cropped patch X needs to adaptively adjusted in order to fit the size of local volumes. And we set $\left\{S_{H}, S_{W}, S_{D}\right\}$ on IRCADb to $\left\{4, 4, 4\right\}$.

Following the baseline \cite{cao2021swin}, we present two successive transformer blocks. The main difference is that our computational unit is built for 3D volumes rather than 2D windows. Based on above volume partitioning way, continues swin transformer can be formulated as follows:

\begin{equation}
\begin{gathered}
\hat{\mathcal T}^{l}=B-M S A\left(L N\left({\mathcal T}^{l-1}\right)\right)+{\mathcal T}^{l-1} \\
{\mathcal T}^{l}=M L P\left(L N\left(\hat{{\mathcal T}}^{l}\right)\right)+\hat{{\mathcal T}}^{l} \\
\hat{\mathcal T}^{l+1}=Shifted \ B-M S A\left(L N\left({\mathcal T}^{l}\right)\right)+{\mathcal T}^{l} \\
{\mathcal T}^{l+1}=M L P\left(L N\left(\hat{{\mathcal T}}^{l+1}\right)\right)+\hat{{\mathcal T}}^{l+1}
\end{gathered}
\end{equation}

Here, $l$ expresses the layer number, MLP represents multi layer perceptron. IB-MSA is our bias multi-head attention and it has the 3D shifted version. Here, the computational complexity of IB-MSA on a volume of HxWxD patches is:
\begin{equation}
\Omega(\mathrm{B}-\mathrm{MSA})=4 h w d C^{2}+2 S_{H} S_{W} S_{D} h w d C
\end{equation}

In the original swin-transformer\cite{liu2021swin} and ViT\cite{dosovitskiy2020image}, the complexity of multi-head  self attention(MSA) is :
\begin{equation}
\Omega(\mathrm{MSA})=4 h w d C^{2}+2(h w d)^{2} C
\end{equation}

The shifted window reduces the computational complexity about 56\% for IRCADb dataset. The \textbf{Fig.3} shows the way where shifted IB-MSA reduces computational complexity by using smaller tiny volume $\left(\left\lfloor\frac{S_{H}}{2}\right\rfloor,\left\lfloor\frac{S_{W}}{2}\right\rfloor,\left\lfloor\frac{S_{D}}{2}\right\rfloor\right)$.

\begin{figure*}[!htb]
  \centering
   {\includegraphics[width=0.9\textwidth]{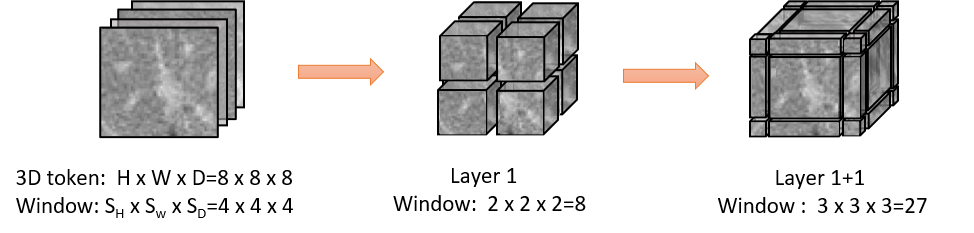}} 
   
  \caption{An illustrated example of $3 \mathrm{D}$ shifted windows. The input size $H^{\prime} \times W^{\prime} \times D^{\prime}$ is $8 \times 8 \times 8$, and the 3D window size $M \times M \times M$ is $4 \times 4 \times 4$. As layer $l$ adopts regular window partitioning, the number of windows in layer $l$ is $2 \times 2 \times 2=8 .$ For layer $l+1$, as the windows are shifted by $\left(\frac{S_{H}}{2}, \frac{S_{W}}{2}, \frac{S_{D}}{2}\right)=(2,2,2)$ tokens, the number of windows becomes $3 \times 3 \times 3=27$. Though the number of windows is increased, the efficient batch computation in [28] for the shifted configuration can be followed, such that the final number of windows for computation is still 8.} 
  
	\label{fig:data_distribution}
	\vspace{0.2in}
\end{figure*}

\hspace*{\fill} \\
\noindent \textbf{IB-MSA and relative position bias matrix}
Different to the first try on ViT\cite{dosovitskiy2020image}, some recent researches\cite{liu2021swin,cao2021swin} have shown that there are a lot advantageous in bias to self-attention computation. 
Here, we intuitively change the bias focus on edge of segmentation volume by introducing 3D relative position bias $B\in \mathbb{R}^{M^{2} \times M^{2} \times M^{2}}$ for each head as:
\begin{equation}
\text { Attention }(Q, K, V)=\text { SoftMax }\left(Q K^{T} / \sqrt{d}+B\right) V
\end{equation}
where $Q, K, V\in \mathbb{R}^{ P \times d}$ are the query, key and value matrices; $d$ is the dimension of query and key features, and $P$ is the number of patch tokens in a $3\mathrm{D}$ window. Since the relative position along each axis lies in the range of $[-2M+1, 2M-1]$ , the positional mask have a big value other than B item. 
we parameterize a smaller-sized bias matrix $\hat{B} \in \mathbb{R}^{(2M-1) \times(2M-1) \times(2M-1)}$, and values in $B$ are taken from $\hat{B}$. 
We know from ViT that, for patch-level embedding inputs, different positional embedding is less important\cite{dosovitskiy2020image}. 2D swin transformer related researches \cite{cao2021swin,liu2021swin} believe the only use relative bias position embedding is better than only use than absolute position embedding. However, relative bias may lose some inductive bias such as locality and translation equivariance. Yet spatial invariance is crucial in our transformer-based design which interleaved with convolution in upsamping/downsamping layers.
Therefore, we adding 1-dimensional absolute positional embedding at the beginning of self-attention computing as inductive bias and then learning to compute new bias matrix.
Our specific setting improved liver vessel edge segmentation in \textbf{Fig. 5} and we observe slight improvement with this bias complement with absolute position. The comparison of other methods is shown in \textbf{Table 2}.     


\subsection{Decoder} \label{decoder set}
In the decoder part, the transformer block are similar to encoder in another direction. Moreover, the up-sampling blocks use deconvolution operator with small kernels and strides which can recover low-level feature to high resolution details quickly if it combined with skip connection. In the final stage, the transformer result combined with local extraction block to output the end-to-end result.   

\subsection{Weighted Loss Function}
Liver vessels only exist in a small region of the liver, and unbalanced foreground(hepatic vessels) and background classes(liver) often cause predictive deviation and bias the classification to the background with more voxels.
It is hard to achieve desired segmentation results with vessels edge and small branch. The similarity matrix of dice coefficient with special penalty weight parameter as $M(P, G, \beta)$ has been proposed to design loss function\cite{huang2018robust} as follows:   
\begin{equation}
M(P, G, \beta)=\frac{|P \cap G|}{|P \cap G|+0.5 \beta(|P-G|+|G-P|)}
\end{equation}
where $\beta$ determined the weight of the number of correctly classified foreground voxels and misclassified voxels.   

Since our task has 2 class labels, we can take foreground and background as the first and second classes, respectively. Then the equation (5) becomes:
\begin{equation}
M(\beta)=\frac{\sum_{i=1}^{N} p_{0 i} g_{0 i}}{\sum_{i=1}^{N} p_{0 i} g_{0 i}+0.5 \beta\left(\sum_{i=1}^{N} p_{0 i} g_{1 i}+\sum_{i=1}^{N} p_{1 i} g_{0 i}\right)}
\end{equation}
where $p_{0 i}$ and $p_{1 i}$ are the probabilities that voxel $i$ belongs to the foreground (liver vessel) and the background (liver), respectively in the softmax layer output result. $g_{0 i}$ and $g_{1 i}$ are the labels of voxel $i$ in the annotated data for liver vessel or liver with value 0 or 1 , respectively.

From \citet{huang2018robust}'s studies, the gradient of similarity in equation(6) to 2 variant shows the weight of the liver(background) and liver vessel(foreground) do not need a pre-trained method unlike \citet{chen2021transunet}, which provided the initial training weights from other models or datasets. Moreover, the proposed algorithms adjust the penalty for misclassified voxels by selecting $\beta$ as 6 can both optimize the dice value and sensitivity in our model.   
 
\subsection{Post-processing}
Due to limitations of the GPU's memory, we cannot put full size volumes into our model. It can cause residual errors in the patch edges.
Therefore, connected component analysis is performed on the vessels after trained by IBIMHAV-Net. To remove some noises caused by classification, regions with small partitions(less than 180 m$m^3$)are removed.         

\section{Experiments and results}
\subsection{Datasets augmentation and experiment material}
3Dircadb datasets (https://www.ircad.fr/research/3d-ircadb-01/) are currently available with liver and liver vessel contours suited to our training and evaluation of liver vessel segmentation algorithms. 
The datasets include 20 contrast-enhanced CT volumes with various image resolutions, vessel structures, intensity distributions, and contrast between liver and liver vessels. To keep the accuracy, transform invariance, and robustness of our network, the training set and test set should have clear, abundant hepatic vessel structures with different intensity ranges, and contrast with background and vessels. The liver vessel appearances should be similar in both training and testing datasets, so we deliver some experiments. By observing the voxel numbers and statistical data, The 3DIRCADb includes 6 simple samples and 14 challenging samples. Finally, we choose 16 volumes and 4 volumes as training/test data separately (both include simple and challenging samples) based on hand selection in each experiment.  For the 16 training sets, we have to applied some image amplification methods for increase our training set. For a sample in training case, the fixed rotation set for 60\degree, 270\degree then add random translation from -25 to +25 pixels to get three times datasets as an augmentation strategy. In both the training and testing datasets, the original pixel spacing varied from 0.56mm to 0.87mm, slice thickness varies from 1.25mm to 2mm and slicer varied from 113 to 225.

\par
Our proposed method was implemented using python 3.8 and PyTorch 1.9.0. All experiments were conducted on an Nvidia A6000 GPU with 48GB memories. Input image size after preprocessing is set as 256x256x192. The crop size based on our network is 128x128x96 with overlapping stride 24 in the test result. The batch size is set to 2, and the learning rate was set as 3e-5, as far as the initial work tested\cite{liu2021swin}, swin transformer can hardly converge in the first 20-30 epochs. In the  training process, we set training epoch as 750. The default optimizer with momentum 0.9 and weight decay 2e-3 was used for model back propagation. We employ precision, dice loss, and sensitivity three indexes to evaluate the results. 

\subsection{Experiments}
In this subsection, we compare the proposed model with other state-of-art methods on 3DIRCADb related work. CNN-based methods including UNet\cite{cciccek20163d}, VNet\cite{milletari2016v}, Huang et al\cite{huang2018robust} which is U-net's optimized variant, and also ResUnet\cite{yu2019liver}. Also the improved graph cuts method proposed by sangsefidi\textit{et.al} which is a practical newly improvement for traditional method has good performance in liver vessel segmentation\cite{sangsefidi2018balancing}. 
In addition, there are some methods applied data refinement\cite{huang2018robust} or specific data augmentation strategy like filters\cite{kitrungrotsakul2019vesselnet}, we do not compare with these results.  
 
\hspace*{\fill} \\
\emph{Quantitative Results} 
To compare with other state-of-art methods in an equitable way, we only focus on original volume 3DIRCADb datasets. Our results are reported in the \textbf{Table 1}. From  \textbf{Table 1}, we can see the numerical results on each metric are larger than other methods. Thanks to swin transformer's shifted window and IB-MSA mechanism, our model adopts larger input to catch global relationship and to obtain better segmentation results. Our method exceed other methods in three indexes. 

\hspace*{\fill} \\
\emph{Visualization Results} 
\begin{figure}
  \centering
  \includegraphics[width=1\textwidth]{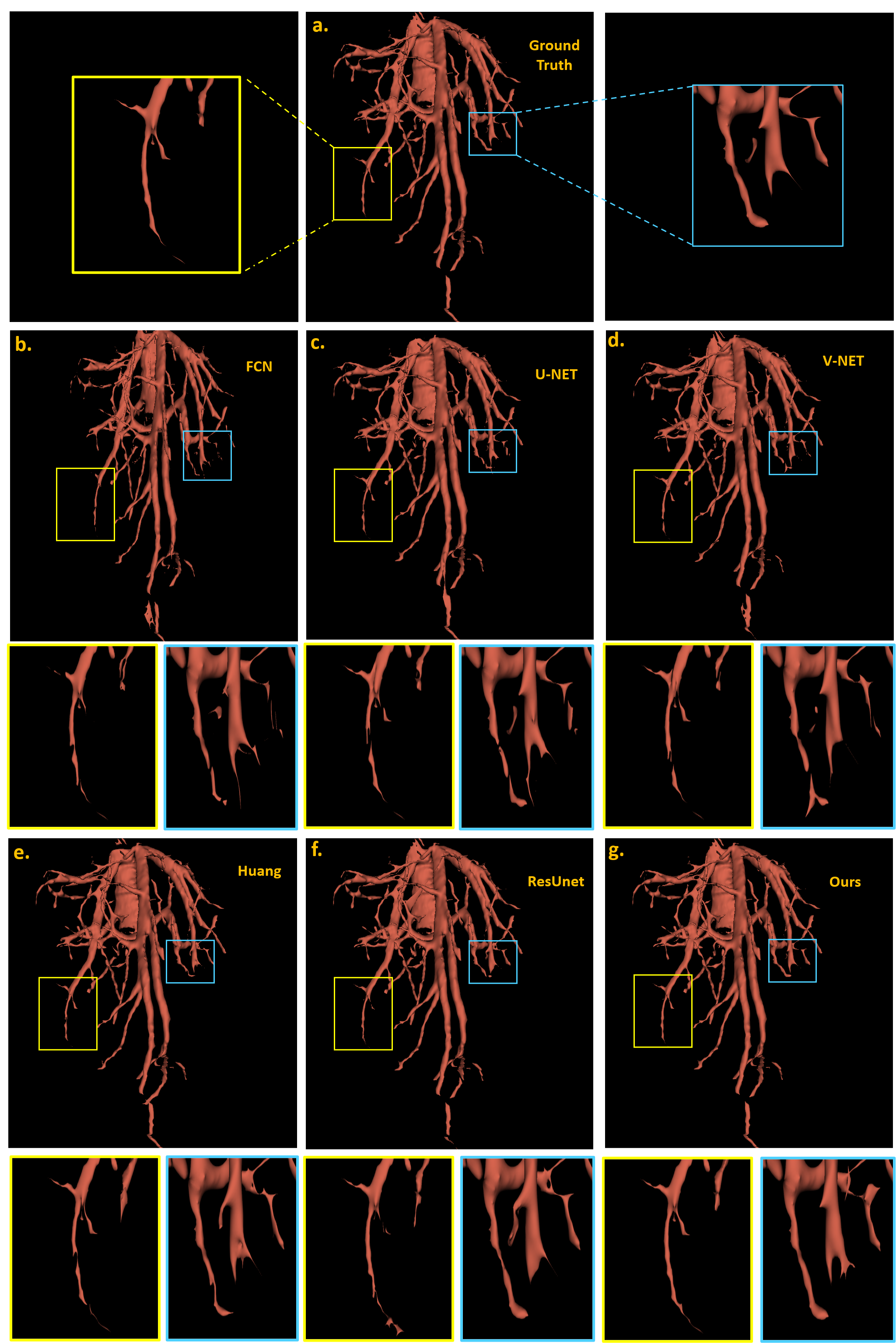}
  \caption{Visualization and comparison of proposed deep learning
method and state-of-art machine learning-based methods using raw volume as in put with post-processing. Three row indicates different genres methods. First row: (a) ground truth result which is most similar to our result. Second row: (b), (c), (d) the traditional 3d medical image methods. Third row:(e), (f), (g) the modern deep learning methods in the journals and our method.}
\end{figure}  
\textbf{Fig. 6} shows the visualization of our experiment in one complex sample. After 3D morphological close
operation and post-processing, the surface of the vessels
becomes smoother and some noise blocks are removed. To compare the results visually, we utilize 3D slicer's toolbox and the zoomed-in patches. The full results are shown below in \textbf{Fig. 6}.  This sample is long and curvy, the segmentation results of FCN and 3D U-Net,3D v-Net on hepatic veins are not so well, in which some regions are over-segmented or some minor vessel are missed. The reason could be Convolutional operators limit the capability of learning long-range dependencies. In addition, the third row's Huang
\textit{et.al} and ResUnet did fairly well in the whole vessel structure, yet have many errors in the vessel edge which can be seen in the zoom-in views. By utilizing the inductive biased multi-head attention and transformer, our methods on vessels performed relatively closer to the ground truth in vessel edges and overall structure.      

\hspace*{\fill} \\
\emph{Evaluation of results} 
\begin{figure*}[!htb]
  \centering
    {\includegraphics[width=0.9\textwidth]{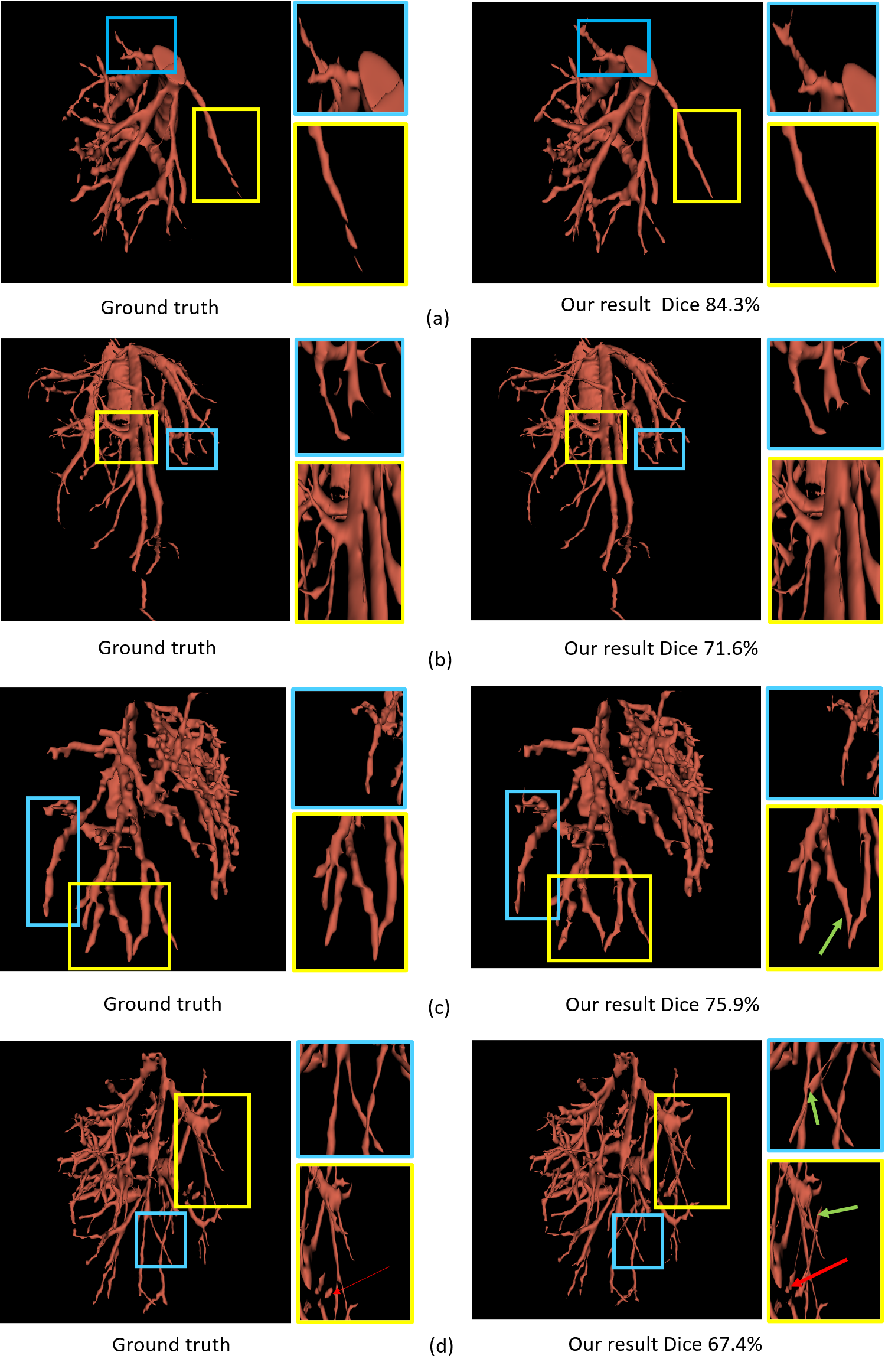}} 
  \caption{The first column list ground truth in different cases. The second column list our network's results (a),(b),(c),(d) represent different cases.}
	\label{fig:IB-MSA result}
	\vspace{0.2in}
\end{figure*}
 To validate the generalization of our method, we conduct 4 test cases with hard cases and simple cases to show the result in \textbf{Table 2}. In simple cases, our network performed very well. As the minor vessel becomes more complex and variant, even if the results have some errors and over-segmentations, the edge segmentation steadily performs different situations. The dice coefficient in this 4 cases is 84.3, 71.6, 75.9, 67.4 respectively. From \textbf{Fig.5} we can see, the simple cases (a)  the vessel edge segmentation may have over-segmentation. In complex cases (c) and (d), the green arrows 
 point to some misclassifications voxels. They are caused by missing labels in the ground truth. The red arrow points to the discontinuous vessel net. It is caused by tumor in that position. In specific, the tumor effects and unlabeled liver vessels in the expert manual annotation both lower the segmentation accuracy.
 
\hspace*{\fill} \\
\renewcommand{\arraystretch}{1.5} 
\begin{table}[tp]  
  
  \centering  
  \fontsize{6.5}{8}\selectfont  
  \begin{threeparttable}  
  \caption{Qualitative comparison of segmentation performance by three evaluation metrics on 3DIRCADb.}  
  \label{tab:performance_comparison}  
    \begin{tabular}{ccccccc}  
    \toprule  
    \multirow{2}{*}{Method}&  
    \multicolumn{3}{c}{G}\cr
    \cmidrule(lr){2-4} 
    &Precision($\%$)&Sensitive($\%$)&Dice($\%$)\cr
    \midrule  
    FCN&80.6${\pm}$15.3&73.8${\pm}$14.2&63.1${\pm}$15.5\cr
    
    UNet\cite{milletari2016v}&87.6${\pm}$11.8&75.8${\pm}$8.4&65.5${\pm}$15.4\cr
    VNet\cite{milletari2016v}&82.1${\pm}$16.7&70.3${\pm}$6.6&64.1${\pm}$11.6\cr
    Huang\textit{et.al}\cite{huang2018robust}&97.1${\pm}$0.8&74.3${\pm}$10.6&67.5${\pm}$6.9\cr
    ResUnet\cite{yu2019liver}&92.6${\pm}$1.4&71.9${\pm}$7.2&70.6${\pm}$8.5\cr
    Graph cuts (Sangse\textit{et.al})\cite{sangsefidi2018balancing}&None&None&74.1${\pm}$12\cr
    IBIMHAV-Net&{\bf 98.8}${\pm}${\bf0.3}&{\bf 78.1}${\pm}${\bf2.4}&{\bf 74.8${\pm}${\bf9.5}}\cr
    \bottomrule  
    \end{tabular}  
    \end{threeparttable}  
\end{table} 

\subsection{Ablation studies} \label{Section:compare}
To explore the influence of our design on the model       
performance,we conducted a series ablation studies on 3Dircadb dataset.

\hspace*{\fill} \\
\noindent \textbf{Influences of inductive biased positional embedding and IB-MSA}
\textbf{Table 2} shows the comparison of different position embedding approaches for our network. IBIMHAV-Net with general position relative bias yields 2.5\% accuracy improvement compared to absolute position embedding, indicating the effectiveness of relative position bias. In addition, our proposed biased attention yields the result better than other positional embedding approaches.

\begin{table}[!ht]
 \caption{Relative position bias}\label{tabl1}
\label{bias table}
\centering
\setlength{\tabcolsep}{3mm}{
\begin{tabular}{c|ccccc} \toprule 
Position embedding methods  &  Precision    \\ \hline
Absolute position embedding  &  95.2       \\
Relative bias embedding  &  97.7    \\
Our relative bias   &  98.8     \\
\bottomrule
\end{tabular}}
\end{table}

\begin{figure*}[!htb]
  \centering
    {\includegraphics[width=0.9\textwidth]{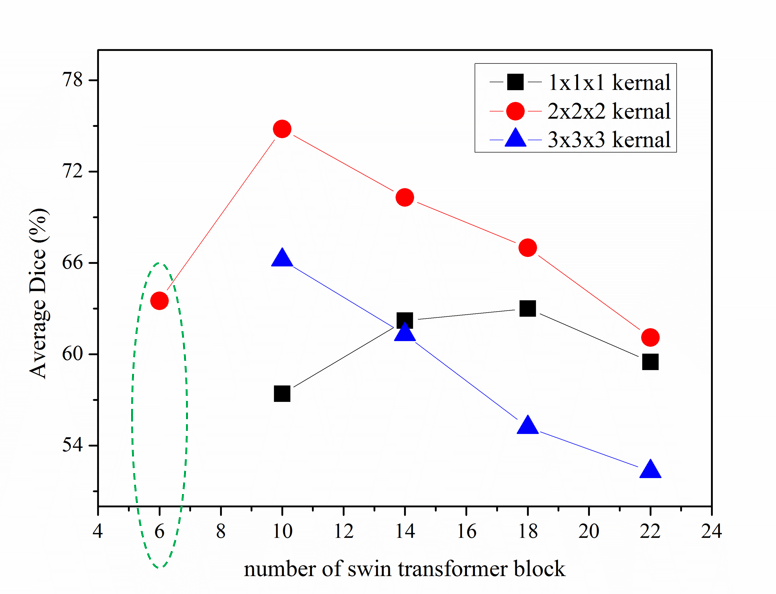}} 
  \caption{The different settings to study effect of kernel size and model bottleneck}
	\label{fig:IB-MSA result}
	\vspace{0.2in}
\end{figure*}

\hspace*{\fill} \\
\noindent \textbf{Influences of more skip conncections and 3D swin-transformer block(bottleneck)}
In our network architecture, the skip connections are connected between after the down-sampling block and before the up-sampling block to unify the feature dimensions. Because the transformer has a different convergence rule compared to CNNs, which need more discussion\cite{cao2021swin,touvron2021going}. In our model, there are Only two successive Swin-transformer blocks are used to learn deep feature representation. Our experiment can only set 6, 10, 14, 18, 22 transformer blocks and corresponding upsampling/downsampling layers to study the convergence pattern of this model, which are shown in \textbf{Fig.7}. It is worth noting that when the number of transformer blocks is 6, the smaller and larger up/down-sampling kernel blocks cannot lead to converge. In 3 groups of ablation experiences, the 2x2x2 kernel and 10 block group gains the best performance for our model.

\hspace*{\fill} \\
\noindent \textbf{Effect of downsampling strategies}
Patch merging is the down sampling strategy  used in original swin transformer and the main idea is concatenetes the neghboring patches\cite{liu2021swin,cao2021swin}. We expand it to 3D by concatenates  2x2x2 neighboring patches first. Then applying a linear layer on the features can downsampling to 2x the original dimension. We choose small kernel convolution layers to reach the same operate and has better results. The results are shown in \textbf{Table.3}
\begin{table}[!htb]
\caption {Ablation study on the impact of down-sampling DSC($\%$)}
\setlength{\tabcolsep}{3mm}{
\begin{tabular}{c|ccccc} \toprule
Down-sampling method  &  DSC for vessel  &  DSC for background    \\ \hline
3D patch merging  & 69.12  & 89.95       \\
convolution with small stride & 74.83  & 96.92     \\
\bottomrule
\end{tabular}}
\end{table}

\hspace*{\fill} \\
\noindent \textbf{Effect of up-sampling strategies}
Original swin transformer choose the patch expanding layer in the encoder based on resize\cite{cao2021swin}, which relies on resizing the patches to upsampling the featueres. We design a small kernel transposed convolution layer in the decoder to perform up-sampling as the feature dimension increases. To explore the proposed strategy effectiveness, we conduct the experiments of IBIMHAV-Net with Trilinear interpolation, 3D transposed convolution and patch expanding layer(\cite{liu2021swin}) on IRCADb datasets. The experimental results are shown in \textbf{Table.4} indicate that the proposed model combined with the transposed convolution layer can obtain better segmentation accuracy.
\begin{table}[!htb]
\caption {Ablation study on the impact of upsampling}
\setlength{\tabcolsep}{3mm}{
\begin{tabular}{c|ccccc} \toprule
Up-sampling method  &  DSC for vessel  &  DSC for background    \\ \hline
patch expanding  & 67.11  & 92.34       \\
trilinear interpolation  &  72.21 & 95.22    \\
3D transposed convolution  & 74.83  & 96.92     \\
\bottomrule
\end{tabular}}
\end{table}

\hspace*{\fill} \\
\noindent \textbf{Effect of local feature extraction block}
The local feature extraction block includes some large kernel convolution layers. We have tried adding other feature extraction residual blocks in deeper places. The experiment shows that the CNNs can only perform well in high resolution part. The main reason may be that the CNNs do not have enough spatial invariant property, which can supplement precise local features for another swin-transformer path. When we dropped this design, the accuracy and Dice coefficient reduced by 12\% and 7.5\%, respectively.

\hspace*{\fill} \\
\noindent \textbf{Rolling of cropping patch size}
The testing results of the proposed IBIMHAV-Net with
224 x 224 x 96, 128 x 128 x 96 input resolutions as input are presented in Table. 3. As the input size increases from 224x224x96 to  and the patch size remains the same as 2, the input token sequence of transformer will become larger, thus improving the segmentation performance of the model. However, although the segmentation accuracy of the model has been slightly improved  $\pm0.3\%$ DSC, the computational load of the whole network has also increased significantly. In order to balance the running efficiency of the algorithm, the experiments in this paper are based on 128x128x96 resolution scale as the input.


\section{Conclusions}
This paper designs a liver vessels segmentation method from CT images using the transformer-based network. Swin transformer has been expanding to 3D as the backbone which interleaved with convolutions and expanding for 3D volumes.  In specific, the small stride convolution in both local feature block path and up/down-sampling blocks keep the spatial information hierarchically for two successive swin transformer blocks. A new pixel-wised embedding method has been used for our few samples task with variant structures. A new type of bias positional embedding in our transformer is proposed. Numerical Evaluation and visualization based on different benchmarks proved the validity of this deep learning method. Our method has been trained and tested on 3DIRCADb datasets. 
In the future, we would further improve segmentation accuracy by introducing more precise datesets and trying multi-task method to reduce negative effects of liver tumors.
\\
\hspace*{\fill} \\
\noindent \textbf{Acknowledges}
\\
\hspace*{\fill} \\
This work is supported by Key-Area Research and Development Program of Guangdong Province, China under Grant 2020B010165004 and National Natural Science Foundation of China with Project No. U1813204.



%
%

\bibliographystyle{plainnat}
\bibliography{refer}
%
%
%
\end{document}